# Traceable instruments for Encircled Angular Flux measurements

Natascia Castagna*[a], Jacques Morel[a], Edward Robinson[b], Hui Yang[b]

[a]Federal Institute of Metrology METAS, Lindenweg 50, 3003 Bern-Wabern, Switzerland; [b]Arden Photonics Ltd, Royston House, 267 Cranmore Boulevard, Solihull B90 4QT, United Kingdom

## ABSTRACT

We report on the development of an instrument for the measurement of the Encircled Angular Flux (EAF) and on establishing its metrological traceability at the required level of uncertainty. We designed and built for that purpose two independent EAF measuring instruments, both based on the analysis of the two-dimensional far field intensity profile observed at the output of an optical fibre, using either CMOS or CCD cameras. An in depth evaluation of the factors influencing the accuracy of the measurements was performed and allowed determining an uncertainty budget for EAF measurements, which was validated by a first series of inter-comparisons. Theses comparisons were performed between the two independent EAF measuring systems, using a 850 nm LED coupled into a gradient index fibre as a test object. We demonstrated a very good equivalence between the two systems, well within the absolute measurement uncertainties that were estimated at the $10^{-3}$ level. Further inter-comparisons using light sources coupled to step-index, large core and small core multimode fibres are still ongoing, with the aim to confirm the performances of the instrument under various illuminating conditions.

## 1. INTRODUCTION

Guided modes in a multimode fibre experience different path lengths and losses. Correct measurements of quantities like insertion loss or bandwidth strongly depend on how different modes are populated. Several metrics for the modal distribution like the Encircled Flux (EF) already exist and are fully traceable, but are not applicable to many type of fibres like step index large core or plastic optical fibres. The EAF was already proposed as a very promising candidate to overcome this issue[1,2]. This metric is based on the evaluation of the far field intensity pattern observed at the output of the fibre, and also allows taking into account possible radial asymmetries in the modal distribution pattern. The sketch in Figure 1 is helpful to understand the measurement principle.

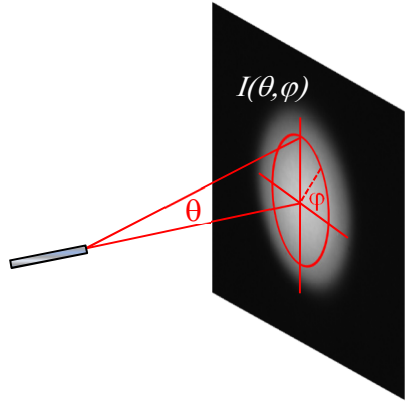

Figure 1. The EAF is a function of the angle $\theta$ and corresponds to the normalised power level observed in a circle defined by each angle $\theta$. Possible asymmetries of the far field power distribution are taken into account by considering the radial power distribution $I(\theta,\varphi)$ of the far-field intensity distribution in the calculation of the EAF.

*natascia.castagna@metas.ch; phone 0041 583870643; www.metas.ch

The EAF, in a polar coordinate frame centred on the profile centroid, is defined as:

$$EAF(\theta) = \frac{\int_0^{2\pi}\int_0^{\theta} I(\theta',\varphi) \frac{\sin(\theta')}{\cos^3(\theta')} d\theta' d\varphi}{\int_0^{2\pi}\int_0^{\theta_{max}} I(\theta',\varphi) \frac{\sin(\theta')}{\cos^3(\theta')} d\theta' d\varphi}, \qquad (1)$$

where $I(\theta,\varphi)$ is the far-field intensity measured at the position $(\theta,\varphi)$ and $\theta$ and $\varphi$ are the polar and the azimuthal angles respectively. $\theta_{max}$ is the angle beyond which no more light is collected and is a function of the numerical aperture $NA$ of the fibre: $NA < n_{air} \sin\theta_{max}$, $n_{air}$ being the refractive index of the air.

We designed and built two independent measuring systems, the first one based on a CCD and the second one on a CMOS camera. The two setups being essentially similar, we focus in the next paragraphs on the description of the second measuring system that is minded to be the EAF reference instrument. We present then in section 3 a first series of inter-comparisons performed between the two systems, using a 850 nm LED coupled into a gradient index fibre as a test object. This allowed demonstrating a very good equivalence between the two systems within the measurement uncertainties that has been demonstrated to be at the $10^{-3}$ level (see section 2).

## 2. EXPERIMENTAL SETUP, TRACEABILITY AND UNCERTAINTY BUDGET

The experimental setup is shown in Figure 2. The fibre under test was placed in front of the camera using a $XY\theta\varphi$ stage to fine tune the position of the fibre and to make sure that the optical axis of the fibre is perpendicular to the camera chip. The alignment was made by optimising the coupling of the light back-reflected by the camera front-window into the fibre. We used for these measurements a 2048 x 2048 pixels, 16 bits, CMOS camera with a pixel size of 6.5 μm.

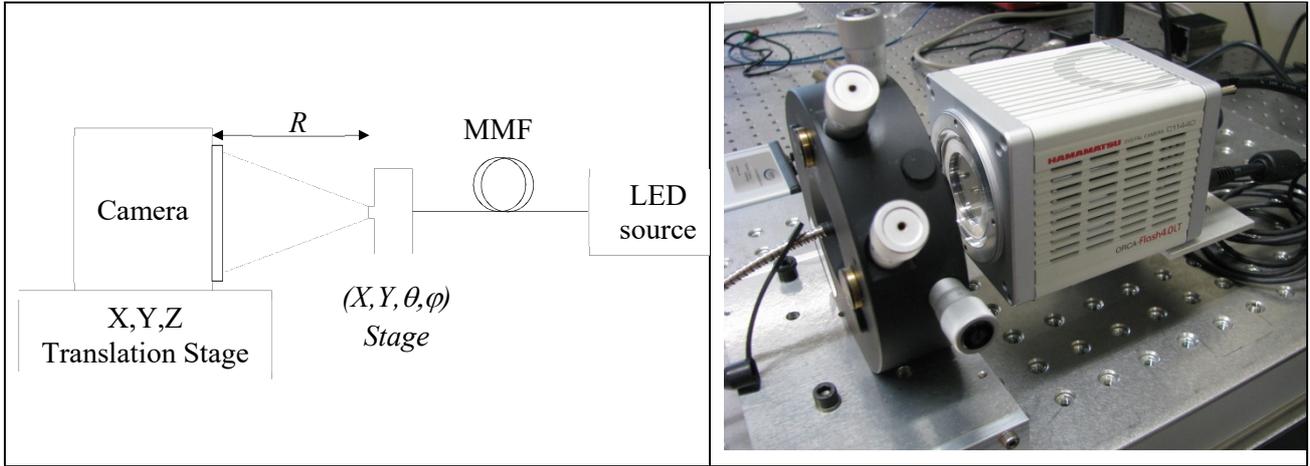

Figure 2. Setup for the measurement of the EAF: a CMOS camera collects the far-field intensity profile at the output of the multimode fibre under test (MMF), which is placed in front of the camera at a fixed known distance $R$.

To establish the traceability of the instrument, all the relevant factors of influence need to be fully identified and quantified. This includes a full evaluation of the optical properties of the camera, like uniformity and linearity and also the analysis of the most critical dimensional and optical parameters of the system. The details of these evaluations are discussed in the next paragraphs, together with a presentation of the overall measurement uncertainty budget.

## 2.1 Uniformity

The uniformity of the CMOS camera was calibrated by comparison to a uniform reference light source which consisted in an integrating sphere illuminated with a series of tungsten lamps. For that purpose, the camera was placed in front of the source and the intensity measured by each camera pixel was recorded. The non-uniformity was then calculated as the difference between the power measured by each pixel and the averaged power level observed by all pixels. The worst case non-uniformity measured on the camera was of 4%. To translate this non-uniformity into an uncertainty on the EAF measurement, we calculated the $EAF_{meas}$ based on the measurement of the uniform intensity profile and compared it to the $EAF_{unif}$ of a theoretical perfectly uniform light distribution and by calculating the difference $\Delta EAF_{Unif} = EAF_{meas} - EAF_{unif}$. The results are given in Figure 3 and showed a maximal deviation $\Delta EAF_{Unif} = 0.0012$ at $\theta = 9°$.

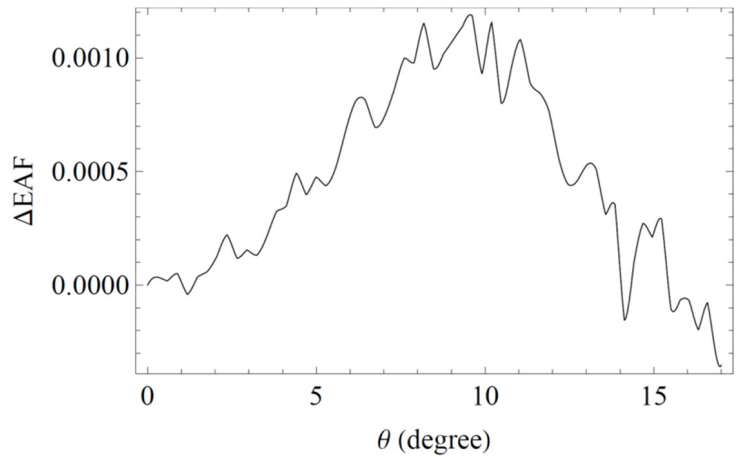

Figure 3. Absolute difference $\Delta EAF_{Unif} = EAF_{meas} - EAF_{unif}$ as a function of the angle $\theta$.

## 2.2 Linearity

The camera linearity was calibrated by comparison to a linearity standard, which consisted in a cooled Ge detector, whose linearity was previously calibrated using a superposition method. The non-linearity of the camera $\eta_{NL}$ was then calculated according to

$$\eta_{NL} = \frac{P_{DUT} - P_{fit}}{P_{fit}}, \qquad (2)$$

where $P_{DUT}$ was the power measured by the camera and $P_{fit}$ the power level calculated from the linear fit of the measured data. For an exposure time of 10 ms and for a light power level ranging from 0.25 to 20 µW, we found $\eta_{NL} \leq 0.5\%$. This power range was chosen to guarantee the best camera linearity conditions. The measurement results of the camera non-linearity as well as the non-linearity coefficient $\eta_{NL}$ are shown in Figure 4 (left). Note that the non-linearity at low power levels can be reduced by increasing the exposure time.

The impact of the residual camera non-linearity on EAF measurements was evaluated by considering a far-field intensity profile $I(\theta,\varphi)$ and by applying to this profile a saturation-induced non-linearity, which was modelled according to

$$I_{NL}(\theta,\varphi) = I(\theta,\varphi) \cdot (1 - \eta_{NL} \frac{I(\theta,\varphi)}{I_{MAX}}), \qquad (3)$$

$I_{MAX}$ being the peak value of $I(\theta,\varphi)$ and $\eta_{NL}$ the worst-case non-linearity of the camera. The EAF deviation $\Delta EAF$ was then calculated according to $\Delta EAF_{Lin} = EAF_{meas} - EAF_{NL}$, where $EAF_{meas}$ was calculated from the initial intensity profile and $EAF_{NL}$ from the saturated profile.

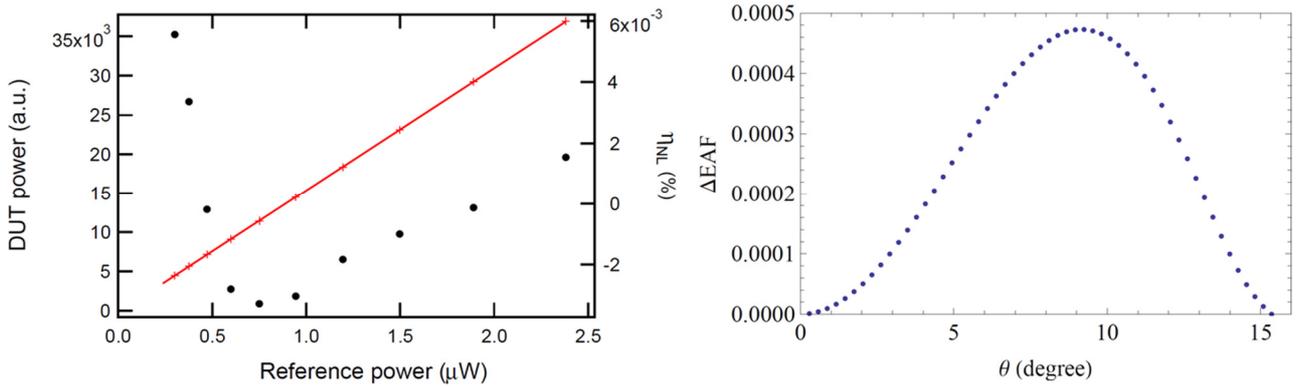

Figure 4. Left: In red, the power level measured by the CMOS camera as a function of the incident reference power. The continuous line is the linear fit of the measured data. The black dots show the non-linear coefficient $\eta_{NL}$ calculated by using Eq.(2). Right: Absolute difference $\Delta EAF_{Lin}$ of the EAF induced by the non-linearity.

This led to a maximum difference of $\Delta EAF_{Lin} = 0.0005$ at $\theta = 9°$, as shown in Figure 4 (right). Note that the difference is the largest at an angle $\theta$ corresponding to the inflection point of the EAF curve.

## 2.3 Dimensional aspects

The distance $R$ between the fibre end-face and the camera surface was calibrated using a reference Optical Low Coherence Reflectometer (OLCR). This quantity plays a crucial role in the computation of the angles $\theta$. The measured distance was of $R = (20320 \pm 0.020)$ mm. The effect of a distance error $\Delta R$ in the determination of the EAF was estimated by calculating the deviation $\Delta EAF_{\Delta R} = EAF_{meas} - EAF_{R+\Delta R}$, where $EAF_{meas}$ was the EAF computed under optimum condition and $EAF_{R+\Delta R}$ was the EAF calculated by considering a distance offset $\Delta R$. The resulting maximum deviation was of $\Delta EAF_{\Delta R} = 0.0015$ at $\theta = 9°$.

Many cameras are protected by a font window whose thickness is large enough to significantly refract the incoming diverging beam and consequently change the size of the measured far field profile. To correct this effect, the window thickness was calibrated with the same reference OLCR and the far-filed intensity distribution was corrected accordingly. The maximum EAF deviation induced by this correction was estimated by simulations and led to a value of $\Delta EAF_W = 0.0013$ at $\theta = 9°$.

The effective pixel size of the camera chip was of 6.5 μm. The uncertainty on EAF measurements induced by the limit of resolution associated to the finite pixel size led to an estimate of the maximum EAF deviation of $\Delta EAF_{\Delta Z} = 0.0002$ at $\theta = 9°$.

## 2.4 Uncertainty budget

The expanded combined standard measurement uncertainty of the EAF measurements was estimated by considering all the above discussed contributions and by including the statistical contribution $uEAF_{Rep}$ arising from repeated measurements. The results are presented in Table 1 for an angle $\theta = 9°$ and were calculated using a coverage factor k=2.

Table 1. Expanded combined standard uncertainty (k=2) of the characterised EAF setup. For the specific fibre that was measured it corresponds to the uncertainty calculated at θ=9°, angle for which the EAF curve has an inflection point.

| Factor of influence | Variable | Value | Distribution | Standard uncertainty |
|---|---|---|---|---|
| Uniformity of CCD camera | $\Delta EAF_{Unif}$ | 0.0012 | rectangular | 0.0007 |
| Non-linearity of CCD camera | $\Delta EAF_{Lin}$ | 0.0005 | rectangular | 0.0003 |
| Fibre to camera distance | $\Delta EAF_{\Delta R}$ | 0.0015 | rectangular | 0.0009 |
| Correction of the refraction through camera window | $\Delta EAF_W$ | 0.0013 | rectangular | 0.0008 |
| Pixel size and limit of resolution | $\Delta EAF_{\Delta Z}$ | 0.0002 | rectangular | 0.0001 |
| Repeatability | $uEAF_{Rep}$ | 0.0001 | normal | 0.0001 |
| **Expanded combined standard uncertainty (k=2)** | | | | **0.0014** |

## 3. INTER-COMPARISONS

A first series of inter-comparison measurements were performed to validate the two setups and their respective analysis tools. A 850 nm LED source coupled into a gradient index fibre, similar to the one used to establish the uncertainty budget, was the test object which was measured with both the METAS and the Arden EAF measuring systems. The measurements results are shown in Figure 5 (left). The EAF curves measured with the two systems are very well comparable, with a maximum EAF deviation of 0.0016, as shown in Figure 5 (right). Supplementary comparisons performed by varying the measuring conditions, such as the distance between the camera and the fibre end or the power level and the camera gains led to comparable results, within the measurement uncertainty.

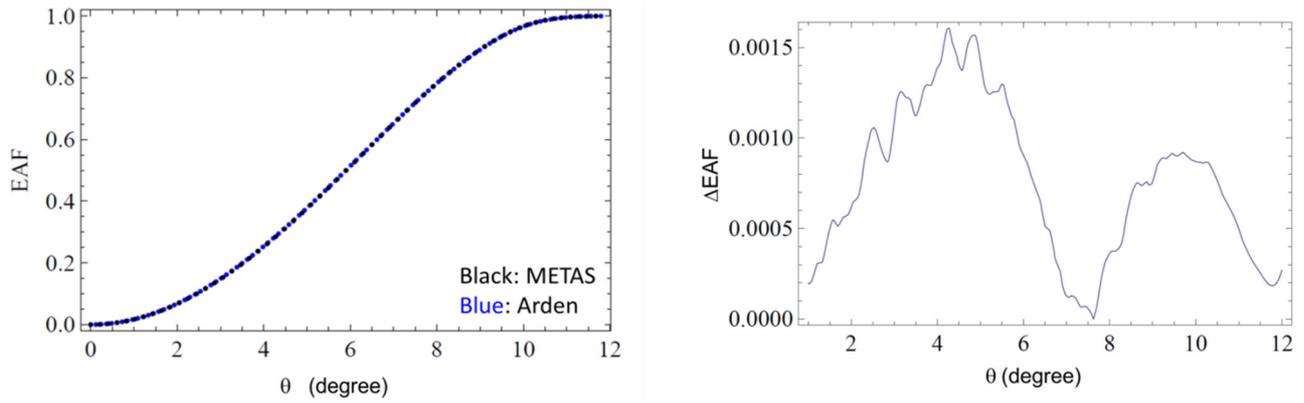

Figure 5 (left) EAF measurements of the 850 nm LED source performed with the METAS system (black curve) and with the Arden system (blue curve). Both measurements are well comparable, within the measurement uncertainty, as shown in Figure 5 (right).

.

## 4. CONCLUSION

The results of this first inter-comparison demonstrate the possibility to perform very well comparable EAF measurements even when using different instruments, which is a prerequisite for a real application of the EAF metrics in the metrology of step-index multimode fibre components and systems, as well as for the definition of EAF reference values in the normative domain.

Further works are still ongoing to validate these two instruments, by measuring different types of step-index multimode fibres and light sources, with the aim to contribute to the normalisation effort in that field.

## 5. ACKNOWLEDGMENTS

This project has received funding from the EMPIR programme co-financed by the Participating States and from the European Union's Horizon 2020 research and innovation programme under grant agreement number 14IND13 (PhotInd).

## REFERENCES

[1] Kagami, M., Kawasaki, A., Yonemura, A, Nakai,M., Mena, P.V. and Selviah, D.R., "Encircled Angular Flux Representation of the Modal Power Distribution and Its Behavior in a Step Index Multimode Fiber", J. Lightw. Technol. 34(3), 943-951 (2016).
[2] Kobayashi, S. and Sugihara, O., "Encircled Angular Flux: A New Measurement Metric for Radiating Modal Power Distributions from Step-Index Multimode Fibers", J. Lightw. Technol. 34(16), 3803-3810 (2016).